\documentstyle[preprint,aps]{revtex}

\begin{document}
\draft
\title{ Quantitative Description of $V_2O_3$ by the Hubbard Model in Infinite
Dimensions}
\author{Jongbae Hong and Hae-Young Kee}
\address{Department of Physics Education, Seoul National University,
Seoul 151-742, Korea}
\maketitle
\begin{abstract}
 We show that the analytic single-particle density of states and the optical
conductivity for the half-filled Hubbard model on the Bethe lattice in
infinite dimensions describe quantitatively the behavior of the gap and the
kinetic energy ratio of the correlated insulator $V_2O_3$. The form of the
optical conductivity shows $\omega^{3/2}$ rising and is quite similar to the
experimental data, and the density of states shows $\omega^{1/2}$ behavior
near the band edges.

\end{abstract}
 It has been an interesting question that how well does the Hubbard model
describe a real strongly  correlated system. To answer this question,
appropriate experiments are required. Recently, Thomas, et. al.\cite{th} have
measured optical conductivity for the insulating state of $V_2O_3$ whose
magnetic phase has been known as antiferromagnetic\cite{mc}. They observed
some interesting features for the properties of the insulating phase of
$V_2O_3$ from the measurement of optical conductivity. They found that the
insulator gap is not the Slater gap come from antiferromagnetic
ordering\cite{sl} but the correlation gap resulted from strong on-site
repulsion\cite{hu}. In other words, the optical conductivity rises as
$(\omega-2\Delta)^{3/2}$, where $2\Delta$ is the gap width, instead of
$\omega^{1/2}$ expected from a Slater antiferromagnet. This $\omega^{3/2}$
rising is expected from the $\omega^{1/2}$ behavior of the single-particle
density of states (DOS) near the band edges. They compared their experimental
data with various theories\onlinecite{ca,ra,gk,inf}. The best fitting one was
the paramagnetic solution of the dynamical mean-field
theory\cite{gk,inf}
which is valid in infinite dimensions. They explained that the reason why
paramagnetic solution describes the insulating state of $V_2O_3$ quite well
was caused by the spin frustration. The spin structure of $V_2O_3$ is
substantially frustrated, i.e., there are both ferromagnetic and
antiferromagnetic nearest neighbors. In addition, it is known that a theory in
infinite dimensions is well-applicable to a bulk system.\cite{fu} They also
argued that their results are representative of the zero-temperature limit.

 In this Letter, we report that the behaviors of $\omega^{3/2}$ in optical
 conductivity and $\omega^{1/2}$ in single-particle DOS can be seen in the
 Hubbard model on the Bethe lattice in infinite dimensions and our results in
 infinite dimensions fit experiments over all the insulator regime of
 $V_2O_3$, while the dynamical mean-field theory shown in Ref. 1 does not
 cover full insulator regime, especially near the metal-insulator transition.
 We show explicitly that the form of the optical conductivity is quite similar
 to that of experiment. In addition, the area under the curve of optical
 conductivity which gives the average kinetic energy has been compared with
 experiment by using the ratios of the averaged kinetic energy to its
 noninteracting value. The agreement between our theoretical values and
 experimental data is quantitatively good.

 We now show our theoretical works and comparison with experiment. Before we
 go further, we first consider spin frustration in the Bethe lattice since
 $V_2O_3$ is considered as a spin frustrated system\cite{th}. We include next
 nearest neighbor hopping to take the frustration into account, since the
 Bethe lattice is a bipartite lattice which does not have frustration in
 lattice arrangement.  For the Bethe lattice, however, the number of next
 nearest neighbor is $q(q-1)$, where $q$ is the coordination number.
 Therefore, $t_2$ representing next nearest neighbor hopping integral must be
 scaled as $t_2=t_*/q$ to make the kinetic energy finite in infinite
 dimensions.  The $t_2$, however,
 is $1/\sqrt{q}$ times less than $t_1$ which represents nearest neighbor
 hopping and scales as $t_1=t_*/\sqrt{2q}$\cite{me}.  Thus the effect
 of next nearest neighbor hopping can be neglected in calculating the DOS on
 the Bethe lattice in infinite dimensions.  Therefore, the result obtained in
 the previous work\cite{hk} can be used without any change for the present
 analysis.

 We briefly introduce the dynamical Lanczos method
 \cite{fu,dago,hong} used in this work.  The single-particle DOS can be
 obtained  by calculating  the one-particle  Green's function of the
 fermion operator at the same site, i.e.,
$\langle\Psi_0|\{c^{\dagger}_{j\sigma}(t),c_{j\sigma}\}|\Psi_0\rangle$
where $c_{j\sigma}^{\dagger}$ and $c_{j\sigma}$ are the fermion creation and
annihilation operators with spin $\sigma$ at site $j$, the curly brackets
mean anticommutator, and $|\Psi_0\rangle$ denotes ground state. The
single-particle DOS $\rho_{\sigma}(\omega)$ is given by\cite{hu}
\begin{equation}
\rho_{\sigma}(\omega)=-\frac{2}{N}\lim_{\epsilon\rightarrow 0^+}\sum_j
{\rm Im} G_{jj}^{(+)}(\omega+i\epsilon),
\end{equation}
where
\begin{eqnarray}
G_{jj}^{(+)}(\omega+i\epsilon)&=     &-\frac{i}{2\pi}
\int_0^{\infty}\langle\Psi_0|\{c_{j\sigma}^{\dagger}(t),c_{j\sigma}\}|\Psi_
0 \rangle e^{i\omega t-\epsilon t}dt \nonumber \\
                              &=     &\frac{1}{2\pi}\langle\Psi_0|
\{c_{j\sigma}, (\omega+L+i\epsilon)^{-1}c_{j\sigma}^{\dagger}\}|\Psi_0\rangle
\nonumber \\
                              &\equiv&-\frac{i}{2\pi}\Xi_{jj}(z)|_{z=-i\omega+
\epsilon},
\end{eqnarray}
where $L$ is the Liouville operator. The superscript $(+)$ denotes the usual
notation of the retarded Green's function\cite{hu}.  The on-site Green's
function $\Xi_{jj}(z)$ can be represented by an infinite continued
fraction\cite{dago,hong},
\begin{equation}
\Xi_{jj}(z)=\frac{1}{z-\alpha_0+\frac{\Delta_1}{z-
\alpha_1+\frac{\Delta_2}{z-\alpha_2+\ddots}}},
\end{equation}
where
$\alpha_{\nu}=(iLf_{\nu},f_{\nu})/(f_{\nu},f_{\nu})$,
$\Delta_{\nu}=(f_{\nu},f_{\nu})/(f_{\nu-1},f_{\nu-1})$.  The  inner product
is defined by $(A,B)=\langle\Psi_0|\{A,B^{\dagger}\}|\Psi_0\rangle$,
where $A$ and $B$ are  operators of the Liouville  space, $B^{\dagger}$ is the
adjoint of $B$. These $\alpha_{\nu}$ and $\Delta_{\nu}$ are obtained using
a recurrence relation $f_{\nu+1}=iLf_{\nu}-
\alpha_{\nu}f_{\nu}+\Delta_{\nu}f_{\nu-1}$.\cite{hong}

The Hubbard model which we study here is written as
\begin{equation}
H=-\sum_{j,l,\sigma}t_{jl}c_{j\sigma}^{\dagger}c_{l\sigma}+\frac{U}{2}\sum_
{j,\sigma}n_{j\sigma}n_{j,-\sigma}.
\end{equation}
By choosing $f_0=c_{j\sigma}^{\dagger}$, we have obtain the on-site Green's
function $\Xi_{jj}(z)$ for the half-filled Hubbard model on the Bethe lattice
in infinite dimensions whose ground state is assumed as paramagnetic,\cite{hk}
\begin{equation}
\Xi_{jj}(\tilde{z})=\frac{\tilde{z}+\frac{\Delta_2}{2b}\left[\frac{(b-a)}
{\tilde{z}}-\tilde{z}\pm\frac{1}{\tilde{z}}\sqrt{(\tilde{z}^2+a-b)^2+4b
\tilde{z}^2}\right]}{\tilde{z}^2+\frac{\Delta_2}{2b}\left[(b-a)-\tilde{z}^2
\pm\sqrt{(\tilde{z}^2+a-b)^2+4b\tilde{z}^2}\right]+\Delta_1},
\end{equation}
where $\Delta_1=\frac{U^2}{4}+\frac{t_*^2}{2}$,
$\Delta_2=(U^2+t_*^2)/(4\Delta_1)$, $a=\frac{U^2}{4}$, $b=1t_*^2$,
and $\tilde{z}=z-i\frac{U}{2}$.

   If  we  set the chemical potential at $\mu=\frac{U}{2}$, Eq. (5) gives the
single-particle DOS for the insulating phase ($a>b$) as follows:
\begin{eqnarray}
\rho_{\sigma}(\omega)&=&\frac{1}{\pi}{\rm
Re} \Xi_{jj}(z)|_{z=-i\omega+0^+} \nonumber \\
                     &=&\frac{\frac{\Delta_1\Delta_2}{2b\pi|\omega|}
                        \sqrt{W}}{\left[\frac{\Delta_2}{2b}(b-a)+\Delta_1
                        +(\frac{\Delta_2}{2b}-1)\omega^2\right]^2+\left[
                        \frac{\Delta_2^2}{4b^2}W\right]}
\end{eqnarray}
where $W=\{\omega^2-(\sqrt{a}-\sqrt{b})^2\}\{(\sqrt{a}+\sqrt{b})^2-
\omega^2\}$. We take $(-)$ sign for $\omega>0$ and $(+)$ for $\omega<0$ to
satisfy
the boundary condition $\Xi_{jj}(t=0)=1$ given in Eq. (2). The lower and upper
Hubbard bands exist $-(\sqrt{a}+\sqrt{b})\leq\omega\leq-\sqrt{a}+\sqrt{b}$ and
$\sqrt{a}-\sqrt{b}\leq\omega\leq\sqrt{a}+\sqrt{b}$, respectively. One can
observe $\omega^{1/2}$ behavior near band edges from Eq. (6).

 One can observe that this single-particle DOS has band width $2D=2\sqrt{b}
 =2t_*$ and band gap $2\Delta=2(\sqrt{a}-\sqrt{b})=U-2D$. Thus we get the
 following relation:
\begin{equation}
\frac{2\Delta}{D}=\frac{U}{D}-2
\end{equation}
We draw Eq. (7) with experimental data in Fig. 1.  A remarkable agreement is
seen over all insulator regime.

 We now obtain the optical conductivity $\sigma(\omega)$ using a formula valid
in infinite dimensions\cite{j},
\begin{equation}
\sigma(\omega)=\sigma_0\int d\omega'\int d\epsilon\rho^{(0)}(\epsilon)\rho(
\epsilon,\omega')\rho(\epsilon,\omega'+\omega)\frac{f(\omega')-
f(\omega'+\omega)}{\omega},
\end{equation}
where $f(\omega)$ is the Fermi distribution function and
$\sigma_0=\frac{\pi t_*^2e^2a^2N}{2\hbar V}$ where $a, N, V$ are lattice
constant, number of lattice sites, volume, respectively.  In Eq.  (8),
$\rho^{(0)}(\epsilon)=\frac{1}{\pi}\sqrt{2t_*^2-\epsilon^2}$ which is the
single-particle DOS for $U=0$, and $\rho(\epsilon,\omega)=-
\frac{1}{\pi}{\rm Im}G(\epsilon,\omega)=-\frac{1}{\pi}{\rm Im}[\omega+i\eta-
\epsilon-\Sigma(\omega)]^{-1}$. Use of the momentum-independence of the
self-energy in infinite dimensions has been made. This property make it
possible to express the self-energy in terms of the on-site Green's function
$G(\omega)=-i\Xi_{jj}(-i\omega)$.

If we set $\zeta=\omega-\Sigma(\omega)$, the one-particle Green's function is
mapped into the frequency renormalized noninteracting one which describes
the noninteracting system under effective field, i.e.,
$G^{(0)}(\zeta,\epsilon)=[\zeta-\epsilon]^{-1}$. Since we obtain
$\alpha_{\nu}=iU/2$, $\Delta_{\nu}=t_*^2/2$ for all $\nu$ for the
noninteracting Hubbard model on the Bethe lattice in the paramagnetic state,
Eq. (3) gives
\begin{equation}
\Xi^{(0)}_{jj}(\tilde{z})=-\left.
\frac{\tilde{z}}{t_*^2}+\frac{\sqrt{\tilde{z}^2+2t_*^2}}{t_*^2}\right
|_{\tilde{
z}=-i\zeta+0^+}=iG^{(0)}(\zeta)=iG(\omega).
\end{equation}
Solving Eq. (9) for $\zeta$ gives the self-energy as
\begin{equation}
\Sigma(\omega)=\omega-\frac{t_*^2}{2}G(\omega)-\frac{1}{G(\omega)}.
\end{equation}
This relation has been obtained by Georges and Krauth\cite{gk} in terms of
the effective action theory.

Using Eqs. (5) and (10), we get $\rho(\epsilon,\omega)$ and finally the
optical conductivity $\sigma(\omega)$ from Eq. (8). Fig. 2 shows theoretical
$\sigma(\omega/t_*)$ in units of $\sigma_0$ for various $U/t_*$. We use
approximation $\Delta_1\approx a$ and $\Delta_2\approx b$ in drawing
$\sigma(\omega/t_*)$. We choose $U/D=2.1$ and $U/D=4$ to compare with
experiment.  To make the comparison appropriate, we need to adjust
$\omega/t_*$ for each sample, since each sample has different band width $D$
which is equal to $t_*$ in our theory.  Therefore, for the horizontal scale,
we set one unit of $\omega/t_*$ to $0.31 {\rm eV}$ for $U/D=4$ and $1.21 {\rm
eV}$ for $U/D=2.1$.
The former gives $2\Delta=0.62{\rm eV}$ and the latter $0.12{\rm eV}$.  Both
gap widths are within experimental error bounds. For the vertical scale,
however, we multiply each $\sigma(\omega)$ by the corresponding $D_m^2$ ($D_m
= 0.47 {\rm eV}$ and $0.31 {\rm eV}$ for $U/D=2.1$ and 4) in Ref. 1 to take
$\sigma_0$ which is proportional to $t_*^2$ into account. Since the vertical
scale itself is arbitrary, only relative height is meaningful. The optical
conductivities represented by experimental scale are shown in Fig. 3 with
$\omega^{3/2}$ rising expressed by the dashed line.

 Even though the theoretical value of the optical conductivity at a particular
frequency is not quite close to that of experiment, overall structure is quite
similar each other. An additional test for our theory related to optical
conductivity can be performed by comparing the ratio of the average kinetic
energy $\langle \hat{T}\rangle$ to its noninteracting counterpart $\langle
\hat{T}\rangle_0$ with experimental data. Since the conductivity sum
rule
\begin{equation}
\int_0^{\infty}\sigma(\omega)d\omega=-\xi\langle\hat{T}\rangle
\end{equation}
gives rise to the average kinetic energy, one can immediately obtain the
average kinetic energy by performing the integration. The explicit form of
the constant $\xi$ has been given by $\xi=\frac{\pi
e^2}{2a\hbar^2}$\cite{kot}. However, we do not need the explicit expression,
because it is cancelled in getting the kinetic energy ratio.

 The noninteracting counterpart $\langle\hat{T}\rangle_0$ can easily be
obtained from Eq. (8) by using $\rho^{(0)}(\epsilon,\omega)=\delta(\omega-
\epsilon)$. Then we get
$\langle\hat{T}\rangle_0=-\frac{\sqrt{2}\sigma_0}{\pi\xi}$.  Using this and
performing the integration in Eq. (11) for the optical conductivities shown in
Fig. 2, we obtain the ratios as follows:
$\langle\hat{T}\rangle/\langle\hat{T}\rangle_0$ = 0.349, 0.205, 0.147, and
0.116 for $U/D$= 2.1, 3, 4, and 5, respectively. We put these values with
experimental data and other theoretical work in Fig. 4. Our theoretical values
are quite close to the result of dynamical mean field theory and its extension
to the metastable branch where both insulating and metallic solutions
coexist\cite{gk,inf}.

 In conclusion, we argue that the paramagnetic solution obtained by the
dynamical Lanczos method for the Hubbard model on a Bethe lattice in infinite
dimensions may describe the insulating phase of the three-dimensional strongly
correlated system like $V_2O_3$ quantitatively well. The spin frustration
effect which makes the system paramagnetic-like has been treated by
considering next nearest neighbor hoppings in the Bethe lattice. The behavior
of the insulating gap and the average kinetic energy ratio according to the
change of $U/D$ is quite well agreed with experiment quantitatively, while the
optical conductivities showing $(\omega-2\Delta)^{3/2}$ rising are agreed
qualitatively. Finally we expect that the metallic regime of $V_2O_3$ can also
be explained quantitatively in terms of our result for metallic phase. This
will be a forthcoming work.

 The authors wish to express their gratitude to the International Center for
Theoretical Physics for financial support and hospitality. They also thank S.
Kim for helpful discussions on numerical computation. One of authors (J.H.)
also thanks the Government of Japan for support. This work has been supported
by SNU-CTP and Basic Science Research Institute Program, Ministry of
Education.

\newpage

{\bf Figure Captions}

\begin{description}
\item[Fig. 1] : Comparison of Eq. (7) with experimental data (solid circles
for optical measurements, open circles and diamonds for dc measurements) of
Ref. 1. The dashed line is the present theory and the solid line is the stable
solution of the dynamical mean field theory.
\item[Fig. 2] : Theoretical optical conductivities for $U/D=2.1, 3, 4$, and 5
obtained by using Eq. (8). We express $\sigma(\omega/t_*)$ in units of
$\sigma_0$.
\item[Fig. 3] : Comparison of optical conductivities for $U/D=2.1$ and 4 with
experimental data (solid points for $U/D\approx 2.1$ and open circles for
$U/D\approx 4$) shown in Ref. 1. The solid lines are theoretical values and
the dashed and the dotted lines denote $\omega^{3/2}$ rising in theory and
experiment, respectively. The horizontal scale for theoretical curves is
reexpressed by the energy scale used in experiment. Arbitrary units are used
for the vertical scale for theoretical values. Details are explained in the
text.
\item[Fig. 4] : Comparison of average kinetic energy ratios
$\langle\hat{T}\rangle/\langle\hat{T}\rangle_0$ for $U/D=2.1, 3, 4$, and 5
with experiment (solid circles) and theory. Present theoretical values are
expressed by solid triangles. The solid, dashed, and dash-dot line denote
dynamical mean field theory given in Ref. 1. The region covered by dashed line
is the metastable regime suggested in Refs. 7 and 8.

\end{description}

\end{document}